\documentclass{aastex}          
\usepackage{spr-astr-addons}    

\begin{document}
%
\title{ Cosmological models coupled with dark matter in a dissipative universe }

\shorttitle{Dark energy and dark matter in dissipative universe}
\shortauthors{Brevik et al.}

\author{I. Brevik\altaffilmark{}}
\email{iver.h.brevik@ntnu.no}
 \affil{Department of Energy and Process Engineering, Norwegian University of Science and Technology, N-7491 Trondheim, Norway}

\and
\author{V. V. Obukhov\altaffilmark{}}
 \affil{Tomsk State Pedagogical University, 634061 Tomsk, Russia, National Research Tomsk State University, Lenin Avenue, 36, 634050 Tomsk, Russia}
 \and
\author{A. V. Timoshkin\altaffilmark{}}
\affil{Tomsk State Pedagogical University, 634061 Tomsk, Russia, National Research Tomsk State University, Lenin Avenue, 36, 634050 Tomsk, Russia}
\email{} 

\altaffiltext{1}{Affilation}

\today

\begin{abstract}

We consider the cosmological system with two interacting fluids: dark energy and dark matter, in a homogeneous and isotropic universe with dissipation. The modified gravitational equation for  dark matter is solved. The analytic representations for the Little Rip, the Pseudo Rip,  and the bounce cosmology models with dissipation are obtained in terms of the thermodynamic parameters in the equation of state. We analyze the corrections in the energy density for  dark matter, in view of  the dissipative processes and the coupling with dark energy.
\end{abstract}


\section{ Introduction}

According to the modern cosmological picture the universe is filled with a negative pressure fluid (dark energy) which accounts for about  73\% of the total mass/energy of the universe and only 27\% of a combination of dark matter and baryonic matter; cf. \cite{riess98,perlmutter99}. One phenomenological approach to describe the dark energy is to assume that this fluid obeys an equation of state with constant thermodynamic parameter $w=p/\rho$, where $\rho$  is the dark energy density and $p$ is the dark pressure; cf. \cite{li11,bamba12,nojiri07}. It is known from  observational data that in the current universe the value of $w$  lies near, or probably below, $-1$, $w=-1.04^{+0.09}_{-0.10}$; cf.  \cite{kowalski08}. The region where $w<-1$ is called the phantom region.

Various cosmological models   for the evolution of the  universe have been presented. They aim at explaining the acceleration era of the universe, and they include future singularity phenomena such as   the dramatic Big Rip [\cite{caldwell03,nojiri04,nojiri03}],  the softer Little Rip [\cite{frampton11,brevik11,frampton12}],  the Pseudo Rip [\cite{frampton12a}],  the Quasi Rip [\cite{wei12}],  and the bounce cosmology [\cite{brandenberger11,novello08,bamba13,cai11,brevik15}].   The Little Rip and the Pseudo Rip phenomena, and the bounce cosmology pictured as an inhomogeneous nonviscous dark fluid coupled with dark matter, were considered by \cite{brevik13,brevik14}. Cosmological models in which the modification of gravity is considered as a viscous fluid are explored by \cite{myrzakul13,myrzakulov14,elizalde14}. Investigations dealing with the coupling between the dark energy and dark matter components are considered by  \cite{brevik14,timoshkin09}. A very extensive review on cosmological evolution with interaction between dark energy and dark matter in general is recently given by \cite{bolotin15}.

 A main motivation for the present study is to investigate the influence from heat {\it dissipation}. We will thus study the Little Rip, the Pseudo Rip, and the bounce cosmology, by considering a homogeneous dark fluid interacting with dark matter. This means that each fluid component constitutes a non-closed physical system, implying exchange of energy with its companion fluid component. Mathematically, it means the appearance of a source term on the right hand side of the energy conservation equation; cf. Eq.~(\ref{1}) below. We will assume that for each fluid component (dark energy and dark matter) there exists a linear and homogeneous equation of state; this being the simplest alternative. We moreover assume a spatially flat, homogeneous and isotropic FRW space.

Analytical expressions  will be  worked out for the cosmological cases mentioned above, and we will make appropriate choices for the various thermodynamical parameters. Energy dissipation is conveniently expressed in terms of the bulk viscosity, as this kind of viscosity is compatible with the assumption about spatial isotropy. It should be mentioned, however, that this is actually a delicate point. The shear viscosity calculated in cosmology on the basis of the Boltzmann equation (cf. \cite{hogeveen86}) is much larger than the bulk viscosity. Thus even a minute anisotropy in the universe might make the shear viscosity an important  component. The shear viscosity will however be omitted here, in accordance with common usage.

Our exposition will be related to the so-called entropic force model - cf. Eq.~(\ref{4}) below - in which the equation of state has quite a general form, expressed in terms of powers of the Hubble parameter.

Finally, as a general point one may ask: what are the physical reasons for introducing viscosity coefficients in the cosmic fluid? This question can be considered from different angles. First, from a hydrodynamical point of view it is apparent that the ideal (nonviscous) fluid model applies only under restricted circumstances. Once there are irreversible process involved, the introduction of one or two viscosity coefficients becomes inevitable. Second, in a cosmological context the inclusion of viscosity broadens the applicability of the theory considerably. Not least so, this applies in connection with the transition to turbulent motion which is quite probable when the fluid approaches the future singularity [\cite{brevik11}]. Even in the very early universe, in connection with the inflationary period, viscosity has been proposed in order to deal with the grateful exit; cf. \cite{myrzakulov14a}. It is thus quite understandable that a large number of papers on viscosity  has appeared on viscous cosmology. Readers interested in review papers may consult, for instance, \cite{gron90,brevik13a}.

\section{	Dissipative cosmologies coupled with dark matter}

We consider a phenomenological model in which there occurs a  {\it constant} entropic pressure associated with the dark matter fluid component. This falls into line with  the so-called entropic cosmology model for a dissipative universe  (see the paper by \cite{komatsu14}  and further references therein). This is a generalized form of cosmology combining bulk viscosity theory with creation theory of cold dark matter.

The modified continuity  and acceleration equations for the interacting fluid system can be written as

\begin{equation}
\left\{ \begin{array}{lll}
\dot{\rho}+3H(\rho +p)=-\tilde{Q}, \\
\dot{\rho}_m+3H(\rho_m +p_e)=\tilde{Q}, \\
\dot{H}=-\frac{k^2}{2}(\rho+p+\rho_m +p_e).
\end{array} \label{1}
\right.
\end{equation}	
Here $H=\dot{a}/a$ is the Hubble parameter, and $k^2=8\pi G$ with Newton's  gravitational constant $G$.  Moreover $\rho, p$ and $\rho_m, p_m$ are the energy density and pressure respectively of dark energy and dark matter, and $\tilde{Q}$ describes the interaction. A dot means derivative with respect to cosmic time $t$. We put $c=1$. Quantities referring to present time $t=0$ are given a subscript zero.

The first line in Eq.~(\ref{1}) is the continuity equation for dark energy [\cite{nojiri05a}]. The second line is the continuity equation for dark matter, in which $p_e$ is the mentioned constant pressure term. This term describes the influence from dissipation, and is connected with the temperature and the entropy of the Hubble horizon.  It describes irreversible entropy, generated either by a bulk viscosity in one  formulation, or with the creation of cold dark matter in the alternative formulation (a closer discussion on this point can be found in \cite{komatsu14} and earlier papers). We assume that the dark matter is dust, so that its mechanical pressure is $p_m=0$.

The constant effective pressure $p_e$ has to be a negative quantity. Compare here with the behavior of viscous cosmology where the thermodynamic pressure $p$ is reduced by an amount $3\zeta H$, where $\zeta$ is the bulk viscosity, necessarily positive from thermodynamic grounds. We shall write it as $p_e \propto -2H_0^2/k^2$, with a positive constant of proportionality. Calling this constant $(\mu -\gamma)$ (because of the use of $\gamma$ in the subsequent Friedmann equation (\ref{3})), we must have $\mu \geq \gamma \geq 0$. Thus
\begin{equation}
p_e=-\frac{2H_0^2}{k^2}\left( \mu-\gamma \right), \label{2}
\end{equation}
and the Friedmann equation is written as
\begin{equation}
H^2=\frac{k^2}{3}(\rho+\rho_m)+\gamma H_0^2. \label{3}
\end{equation}
The additional term $\gamma H_0^2$ corresponds to an entropic-force term derived from reversible entropy in the entropic cosmology model; cf. \cite{komatsu14}. The term can be interpreted as an effective dark energy term. The constant effective pressure $p_e$ in Eq.~(\ref{2}) consists of contributions from the entropy term to the pressure. They turn up in the coefficients $w_i(t)$ given in Eq.~(\ref{4}) below.

We  consider the following general equation of state for the dark energy [\cite{capozziello05,nojiri06}]
\begin{equation}
p=w_4(t)H^4+w_3(t)H^3+w_2(t)H^2+w_1(t)H, \label{4}
\end{equation}
where the thermodynamic parameters $w_1,w_2,w_3,w_4$ are so far undetermined functions of the cosmic time $t$. Higher order terms are neglected because inflation in the early universe is not considered.

We turn now to  dissipative properties of the interacting dark energy universe, via proper adjustments of the parameters in Eq.~(\ref{4}).

\subsection	{Little Rip model}

Let us assume the Little Rip model where the Hubble parameter increases exponentially with $t$,
\begin{equation}
H=H_0\,e^{\lambda t}, \quad H_0>0, ~\lambda >0 \label{5}
\end{equation}
(recall that $t=0$ is present time).

Consider again the continuity equation for dark matter
\begin{equation}
\dot{\rho}_m+3H(\rho_m +p_e)=\tilde{Q}, \label{6}
\end{equation}
and insert the following expression for the interaction term
\begin{equation}
\tilde{Q}=\frac{H}{\lambda}H_0^5. \label{7}
\end{equation}
In this case the solution of Eq.~(\ref{6}) for the dark matter is given by
\begin{equation}
\rho_m(t)=(\rho_0-\Delta \rho_m)\exp \left( -\frac{3(H-H_0)}{\lambda}\right) +\Delta \rho_m, \label{8}
\end{equation}
where $\rho_m(0)=\rho_0$ and the constant correction term $\Delta \rho_m$  is
\begin{equation}
\Delta \rho_m=\frac{H_0^5}{\lambda}-p_e. \label{9}
\end{equation}
From Eq.~(\ref{8}) one sees that $\rho_m(t\rightarrow \infty)=\Delta \rho_m$. This correction is caused by the exchange of energy between dark fluid and dark matter and by dissipative properties in the universe.

We now start from the  gravitational equation of motion
\begin{equation}
\frac{2\lambda H}{k^2}+\rho+p+\rho_m+p_e=0. \label{10}
\end{equation}
 Choosing for the parameter $\gamma$ the value
 \begin{equation}
 \gamma=\frac{p_e}{3}\left( \frac{k}{H_0}\right)^2, \label{11}
 \end{equation}
 and choosing for the parameter $w_1(t)$ in the equation of state (\ref{4}) the value
 \begin{equation}
 w_1(t)=-\frac{2\lambda}{k^2}, \label{12}
 \end{equation}
 we can rewrite Eq.~(\ref{10}) in the simple form
 \begin{equation}
 w_4H^2+w_3H+\left( w_2+\frac{3}{k^2}\right)=0. \label{13}
 \end{equation}
 Thus, if we require that the parameters $w_2, w_3,w_4$ are related via the condition
 \begin{equation}
 w_3^2- 4w_4\left( w_2+\frac{3}{k^2}\right)=0, \label{14}
 \end{equation}
 we can from Eq.~(\ref{13}) determine the thermodynamic parameters as
\begin{equation}
\left\{ \begin{array}{lll}
w_3(t)=-2H_0 w_4(t) \exp(\lambda t), \\
w_2(t)=w_4(t)H_0^2 \exp(2\lambda t)-3/k^2, \\
w_4(t) \in R.
\end{array} \label{15}
\right.
\end{equation}	
Insertion of these values in the equation of state (\ref{4}) shows that the dark fluid pressure $p$ acquires an exponentially increasing behavior when $t\rightarrow \infty$. Thus the values for the $w$'s chosen in Eqs.~(\ref{12}) and (\ref{15}) lead to the characteristic features of the Little Rip cosmology. This circumstance motivates our parameter choices,  including the ansatz (\ref{11}) for $\gamma$. Physically, we have as a checking point that $\tilde{Q} \rightarrow 0$ when $t\rightarrow \infty$.

\subsection{Pseudo Rip model}

Now we will investigate a cosmological model where the dark energy is monotonically increasing. In this case the Hubble parameter asymptotically approaches a cosmological constant, which
means that the universe approaches a de Sitter form.

We suppose that the Hubble parameter is given by [\cite{frampton12}]
\begin{equation}
H=H_0-H_1e^{-\tilde{\lambda}t}, \label{16}
\end{equation}
where $H_0, H_1$ and $\tilde \lambda$ are positive constants, $ H_0 >H_1>0, \,t>0$.

Near the present time ($t\rightarrow 0$) we have $H\rightarrow H_0-H_1$, and in the late-time universe $(t\rightarrow \infty)$ the Hubble parameter tends to a constant.

We take now the coupling term to have the form
\begin{equation}
\tilde{Q}=\delta H (\rho_m+p_e), \label{17}
\end{equation}
where $\delta$ is a positive nondimensional constant. This choice is motivated by the physically natural assumption that the coupling increases when the relative rate of change of the scale factor increases, i.e. when $H$ increases.

From Eq.~(\ref{6}) we find the energy density of dark matter
\begin{equation}
\rho_m(t)=(\rho_0+p_e)\exp \left[(\delta-3)\left( H_0t-\frac{H-H_0+H_1}{\tilde{\lambda}}\right)\right]-p_e, \label{18}
\end{equation}
with $\rho_m(0)=\rho_0$ as before.

The gravitational equation of motion for dark energy becomes
\begin{equation}
\frac{2\tilde \lambda}{k^2}(H_0-H)+p+\frac{3}{k^2}H^2+p_e-\frac{3\gamma}{k^2}H_0^2=0, \label{19}
\end{equation}
with $\tilde{\lambda}=\frac{3}{2}\gamma H_0$, $\gamma$ being given by Eq.~(\ref{11}).

Thus, if we choose the thermodynamic parameters in the equation of state (\ref{4}) to have the form
\begin{equation}
\left\{ \begin{array}{llll}
w_1=\frac{3\gamma}{k^2}H_0, \\
w_2=-\frac{3}{k^2}, \\
w_3(t)=\left[ H_1\exp \left(-\frac{3}{2}\gamma H_0t\right)-H_0\right] w_4(t), \\
w_4(t) \in R,
\end{array} \label{20}
\right.
\end{equation}
we obtain the characteristics of the Pseudo Rip dissipative universe. Assuming that $w_4$ stays finite when $t\rightarrow \infty$, all the other $w$'s stay finite, and prevent the pressure in Eq.~(\ref{4}) from blowing up in the far future. Again, our choices of parameters are justified by the adjustment to one of the future singular cosmological models.

\subsection{Bounce cosmology}

In the matter bounce cosmology model, the universe goes from an era of accelerated collapse into an era of expansion via a bounce acting as a singularity. After the bounce, the universe is in a matter dominated phase.

Let us consider such a model where the scale factor $a$ has the exponential form [\cite{bamba13}]
\begin{equation}
a= \exp(\beta t^2), \label{21}
\end{equation}
$\beta$ being a positive constant.The bounce occurs at $t=0 \,(a=1)$. We assume that $t \ge 0$ in the following.
Now take the Hubble parameter to have the form
\begin{equation}
H=2\beta t, \label{22}
\end{equation}
and take the interaction between dark energy and dark matter to be
\begin{equation}
\tilde{Q}=Q_0\exp(-3\beta t^2), \label{23}
\end{equation}
with $Q_0=\tilde{Q}(0)$. Thus near the bounce $\tilde{Q}\rightarrow Q_0$, whereas for increasing values of $t$, $\tilde{Q}\rightarrow 0$ and the coupling becomes negligible.

Now the solution of Eq.~(\ref{6}) for the dark matter becomes
\begin{equation}
\rho_m(t)=(Q_0t+\rho_0+p_e)\exp(-3\beta t^2)-p_e, \label{24}
\end{equation}
with $\rho_m(0)=\rho_0$ as before. Just after the bounce the dark matter energy density thus increases linearly with $t$, whereas when $t\rightarrow \infty$, $\rho_m \rightarrow -p_e$. The correction $-p_e$ takes into account the dissipative processes.

 We make the following choice for   the parameter $\beta$,
\begin{equation}
\beta=\frac{3}{4}\gamma H_0^2, \label{25}
\end{equation}
and write the gravitational equation for dark energy as
\begin{equation}
\frac{4\beta}{k^2}+\frac{3}{k^2}(H^2-\gamma H_0^2)+p+p_e=0. \label{26}
\end{equation}
The conditions (\ref{4}) and (\ref{25})  lead to the simple form (\ref{26}):
\begin{equation}
w_4H^4+w_3H^3+\left(w_2+\frac{3}{k^2}\right) H^2+w_1H+p_e=0. \label{27}
\end{equation}
From Eq.~(\ref{26}) one can find the following solution
\begin{equation}
\left\{ \begin{array}{llll}
w_1(t)=-\frac{2}{3}\frac{p_e}{\gamma H_0^2}\frac{1}{t}, \\
w_2=-\frac{3}{k^2}, \\
w_3(t)=  -\frac{3}{2}\gamma H_0^2w_4(t)t,  \\
w_4(t) \in R.
\end{array} \label{28}
\right.
\end{equation}
It is seen that the singularity property at the bounce at $t=0$ is  taken care of by the coefficient $w_1$.

We have thus in this subsection presented a description of dissipative bounce cosmology, related to the equation of state (\ref{4}).

\section{	Conclusion}
 We have investigated some examples of  Little Rip, Pseudo Rip, and the bounce cosmology, under conditions where a dark energy fluid is coupled with dark matter via an interaction term $\tilde{Q}$; cf. Eq.~(\ref{1}). The underlying geometry is  a flat Friedmann-Robertson-Walker universe. The energy dissipation  is most naturally described by means of a bulk viscosity, although there is a mathematically equivalent formalism implying creation of cold dark matter.

 The formalism that we have made use of, falls within the so-called entropic force model [\cite{komatsu14}].  We have assumed initially  that   the equation of state for the dark fluid has the general form (\ref{4}), in which there occurs four thermodynamic coefficients $w_i$, $i \in [1,4]$. The various categories of singularity-containing universes are then obtained by appropriate choices of these coefficients. Corrections to the dark matter energy density caused by dissipative processes are obtained.

 \bigskip
 Finally, we wish to make some additional comments on the physical meaning of the constant pressure $p_e$  introduced in Eq.~(\ref{1}),  from the viewpoint  of viscous cosmology. As mentioned above, the modification of thermodynamic pressure in a viscous fluid consists in replacing the thermodynamic pressure $p$ with $ p-3\zeta H$, with $\zeta$  the bulk viscosity. Thus, the quantity $p_e$ corresponds to $-3H\zeta$, and the assumed constancy of $p_e$ implies that $\zeta$ has to be inversely proportional to $H$.

 It is worth noticing that when this condition is satisfied, the Friedmann formalism for the viscous fluid becomes easily manageable. Let us recall from \cite{brevik11} how the formalism looks if one starts from the equation of state in the form
 \begin{equation}
 p=-\rho -A\sqrt{\rho}+p_e \label{29}
 \end{equation}
 for the dark fluid. Here $A$ is a constant, with dimension cm$^{-2}$ in geometric units. In this case we find the following explicit formula for how $\rho$ depends on cosmic time  $t\, (>0)$,
 \begin{equation}
 t=\frac{2}{\sqrt{3}\,k}\frac{1}{A}\ln \frac{A\sqrt{\rho}-p_e}{A\sqrt{\rho_0}-p_e}. \label{30}
 \end{equation}
 This can be inverted to give $\rho$ as a function of $t$,
 \begin{equation}
 \rho(t)=\left[ \left(\sqrt{\rho_0}-\frac{p_e}{A}\right)\exp \left( \frac{\sqrt{3}}{2}kAt\right)+\frac{p_e}{A}\right]^2. \label{31}
 \end{equation}
 It is seen that the universe needs an infinite time to reach the state $\rho \rightarrow \infty$. This is just the characteristic property of Little Rip cosmology.

\section*{Acknowledgments}

This work was supported by a grant from the Russian Ministry of Education and Science, project TSPU-139 (A.V.T. and V.V.O.).

%



\bibliographystyle{spr-mp-nameyear-cnd}

 \bibliography{<bib data>}                

%

\end{document}